\documentclass[a4paper,fleqn,usenatbib]{mnras}

\usepackage{newtxtext,newtxmath}

\usepackage[T1]{fontenc}
\usepackage{ae,aecompl}

\usepackage{graphicx}	
\usepackage{amsmath}	
\usepackage{amssymb}	

\title[The Cause of the Green Valley/Mountain]{The Causes of the Red Sequence, the Blue Cloud, the Green Valley
and the
Green Mountain}

\author[S. A. Eales  et al.]{
Stephen A. Eales,$^{1}$\thanks{E-mail: sae@astro.cf.ac.uk}, Maarten Baes$^2$, Nathan Bourne$^3$, 
Malcolm Bremer$^4$ \newauthor
Michael J.I. Brown$^5$, 
Christopher Clark$^1$, David Clements$^6$, 
Pieter de Vis$^7$, \newauthor
Simon
Driver$^8$, Loretta Dunne$^{1,3}$, Simon Dye$^9$, Cristina Furlanetto$^{10}$, \newauthor 
Benne Holwerda$^{11}$, R.J. Ivison$^{12}$, L.S. Kelvin$^{13}$, 
Maritza Lara-Lopez$^{14}$,\newauthor
Lerothodi Leeuw$^{15}$,
Jon Loveday$^{16}$, 
Steve Maddox$^{1,3}$, 
Micha{\l} J.~Micha{\l}owski$^{17}$,\newauthor
Steven Phillipps$^4$,
Aaron Robotham$^8$, 
Dan Smith$^{18}$, Matthew Smith$^1$,\newauthor 
Elisabetta Valiante$^1$, Paul
van der Werf$^{19}$ 
and
Angus Wright$^{20}$\\
$^{1}$School of Physics and Astronomy, Cardiff University, The Parade, Cardiff CF24 3AA, UK\\
$^{2}$Sterrenkundig Observatorium, Universiteit Gent, Krijgslaan 281 S9, B-9000 Gent, Belgium\\
$^3$Institute for Astronomy, The University of Edinburgh, Royal Observatory, Blackford Hill,
Edinburgh, EH9 3HJ, UK\\
$^4$Astrophysics Group, Department of Physics, University of Bristol, Tyndall Avenue,
Bristol BS8 1TL\\
$^{5}$School of Physics and Astronomy, Monash University, Clayton, Victoria 3800, Australia\\
$^6$ Astrophysics Group, Imperial College London, Blackett Laboratory, Prince Consort Road,
London SW7 2AZ, UK\\
$^7$ Institut d'Astrophysique Spatiale, CNRS, Universit\'e Paris-Sud,
Universit\'e Paris-Saclay, Bat. 121, 91405,
Orsay Cedex, France\\
$^8$International Centre for Radio Astronomy Research, 7 Fairway, The University of Western\\
Australia, Crawley, Perth, WA 6009, Australia\\
$^9$School of Physics and Astronomy, University of Nottingham, University Park, Nottingham
NG7 2RD, UK\\
$^{10}$ Instituto de F\'isica, Universidade Federal do Rio Grande do Sul,
Av. Bento Gon\c{c}alves, 9500, 91501-970, Porto Alegre, RS, Brazil\\
$^{11}$ Department of Physics and Astronomy, 102 Natural Science Building,
University of Louisville, Louisville KY 40292, USA\\
$^{12}$European Southern Observatory, Karl-Schwarzschild-Strasse 2, 85748, Garching, Germany\\
$^{13}$ Astrophysics Research Institute, Liverpool John Moores University, 146 Brownlow Hill, Liverpool
L3 5RF, UK\\
$^{14}$ Dark Cosmology Centre, Niels Bohr Institute, University of Copenhagen,
Juliane Maries Vej 30, DK-2100 Copenhagen, Denmark\\
$^{15}$College of Graduate Studies, University of South Africa, P.O. Box 392, UNISA, 0003, South Africa\\
$^{16}$Astronomy Centre, University of Sussex, Falmer, Brighton BN1 9QH, UK\\
$^{17}$ Astronomical Observatory Institute, Faculty of Physics,
Adam Mickiewicz University, ul.~S{\l}oneczna 36, 60-286 Pozna{\'n}, Poland\\
$^{18}$Centre for Astrophysics Research,
School of Physics, Astronomy and Mathematics, University of Hertfordshire,\\
College Lane, Hatfield, AL10 9AB, UK\\
$^{19}$ Leiden Observatory, PO Box 9513, 2300 RA Leiden, the Netherlands\\
$^{20}$ Argelander-Institut fur Astronomie, Auf dem Hugel 71, D-53121 Bonn,
Germany
}

\date{Accepted XXX. Received YYY; in original form ZZZ}

\pubyear{2015}

\begin{document}
\label{firstpage}
\pagerange{\pageref{firstpage}--\pageref{lastpage}}
\maketitle

\begin{abstract}
The galaxies found 
in optical surveys fall in two distinct regions of a diagram of optical colour versus
absolute magnitude: the red sequence and the blue cloud
with the green valley in between.
We show that the galaxies found in a submillimetre survey have almost the opposite
distribution
in this diagram, forming a `green mountain'. 
We show that
these distinctive distributions follow naturally from 
a single, continuous, curved Galaxy Sequence 
in a diagram of specific star-formation rate versus stellar mass
without there being the need for
a separate star-forming galaxy Main Sequence and region of passive galaxies.
The cause of the red sequence and the blue cloud 
is the geometric mapping between stellar mass/specific star-formation rate and absolute magnitude/colour,
which distorts a continuous Galaxy Sequence in the diagram of intrinsic
properties into a bimodal distribution in the diagram of observed
properties. The cause of the green mountain is Malmquist bias in the
submillimetre waveband, with submillimetre surveys tending to select galaxies
on the curve of the Galaxy Sequence, which have the highest ratios of submillimetre-to-optical
luminosity. This effect, working in reverse, causes galaxies on the curve of the
Galaxy Sequence to be underrepresented in optical samples, deepening the green valley.
The green valley is therefore
not evidence (1) for there being two distinct populations of galaxies, (2) 
for galaxies in this region evolving more quickly than galaxies in the blue cloud and
the red sequence, (c) 
for rapid quenching processes in the
galaxy population.
\end{abstract}

\begin{keywords}
galaxies: evolution
\end{keywords}



\section{Introduction}

Galaxies discovered in
optical surveys fall in two main areas of a diagram of optical colour
versus absolute magnitude: a narrow band of galaxies with red colours,
commonly called the `red sequence', and a more diffuse `blue cloud'. In between these two
regions there are still galaxies, but fewer, and this region is commonly called
the `green valley'. The existence of the red sequence has been known
for over 50 years (Baum 1959), and the form of the rest of the distribution
has
gradually become clear over the intervening decades (e.g. Visvanathan 1981), with a big increase in
our knowledge coming with the release of the huge galaxy catalogues
produced from the Sloan Digital Sky Survey
(SDSS; Strateva et al. 2001; Bell et al. 2003).

The obvious interpretation of this
diagram (we will argue
in this paper that it is not necessarily the correct one) 
is that there are two physically distinct classes of galaxy.
Additional evidence in favour of this interpretation
is that the morphologies of the
galaxies in the two classes are also generally different,
with the galaxies in the red sequence mostly being early-type
galaxies (ETGs) and the blue-cloud galaxies generally being late-types
(LTGs), although the 
correspondence is not perfect and the exceptions have spawned an entire
literature
(Cortese 2012 and references therein).

Once the form of this distribution had become clear, the natural next step
was to go beyond
plotting colour-magnitude
diagrams and instead 
plot real physical properties of galaxies. Since blue colours indicate a galaxy with
a high specific star-formation rate (star-formation rate divided by stellar
mass, SSFR), with red colours indicating the opposite, and absolute magnitude
being approximately proportional to stellar mass, 
a natural diagram to plot was
SSFR versus stellar mass.
Most of the early papers 
concluded that galaxies fall in two distinct regions of this
diagram. This time, because of the mapping between colour and SSFR, the
blue galaxies fell in a narrow band, which was given the name the `Main Sequence'
or the `Star-Forming Main Sequence', while the red sequence became a region
separated from and below the Main Sequence, with the galaxies in this region
being either called `red and dead', `quiescent' or `quenched'   
(Noeske et al. 2007; Daddi et al. 2007; Elbaz et al. 2007; Peng et al. 2010; Rodighiero
et al. 2011).

\begin{figure*}
\includegraphics[width=140mm]{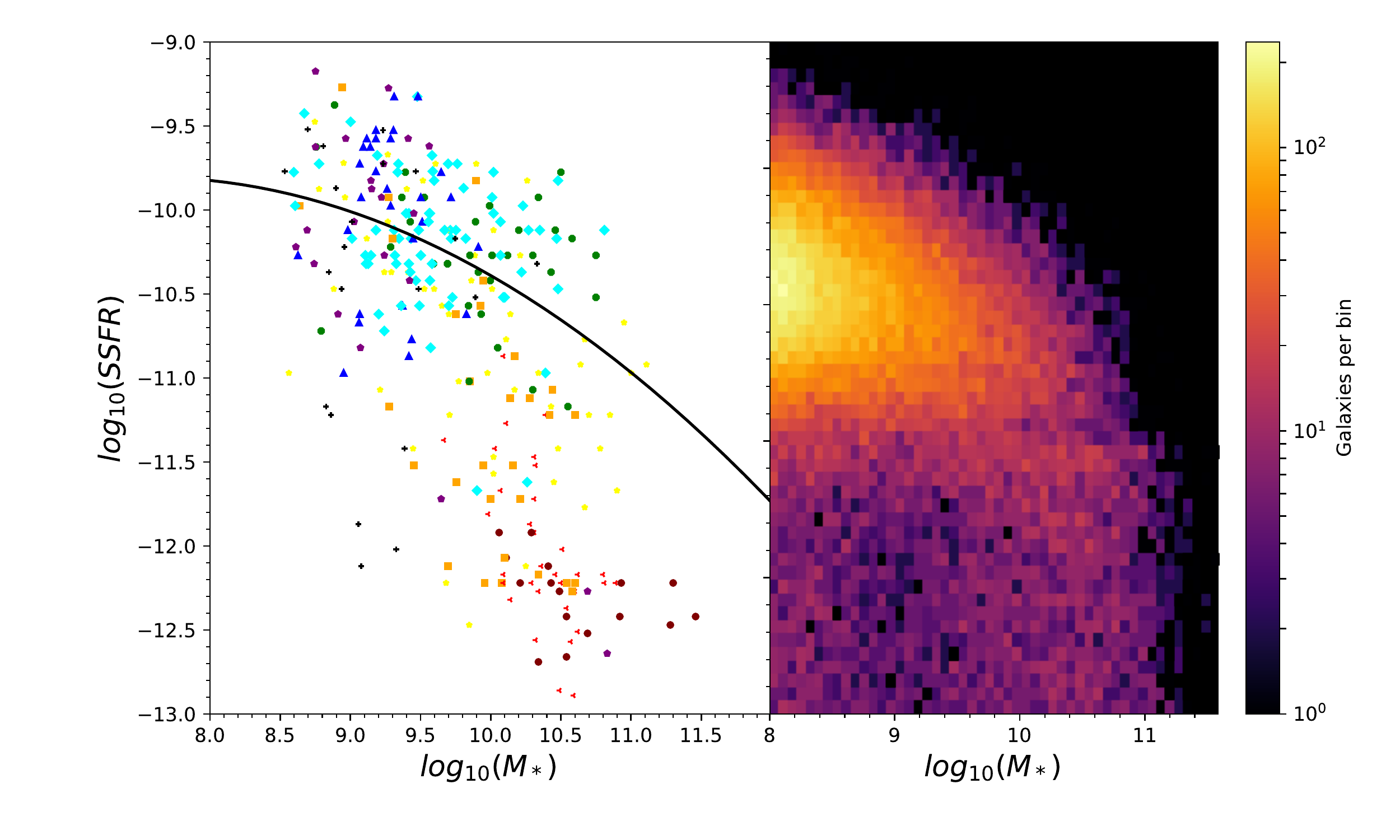}
  \caption{{\bf Left:} Specific star-formation rate versus stellar
mass for the galaxies in the {\it Herschel} Reference Survey (E17).
The values of SSFR and stellar mass are estimates from the
MAGPHYS SED-modelling program (Da Cunha et al. 2008).
The colour and shape of each marker indicate the morphology of the galaxy: maroon circles - E and E/S0;
red left-pointing triangles - S0; orange squares - S0a and Sa; yellow stars - Sab and Sb; green octagons
 - Sbc; cyan diamonds - Sc and Scd;
blue triangles- Sd, Sdm; purple pentagons - I, I0, Sm and Im. The crosses
show galaxies for which there is not a clear morphological
class (Boselli et al. 2010).
The line is
a fit to the HRS galaxies with $log_{10}(SSFR) > -11.5$, 
which is the region of the diagram covered by late-type galaxies.
The line
is used in the Monte-Carlo
simulation described in \S4, and 
has the form $log_{10}(SSFR) = -10.39 - 0.479(log_{10}(M_* - 10.0)
-0.098(log_{10}(M_*) - 10.0)^2$. {\bf Right:} The artificial sample
of 40,000 galaxies generated in the Monte-Carlo simulaton described
in \S4, plotted in the same
diagram as the HRS galaxies. 
}
\end{figure*}

The consequence of there being two physically distinct classes of galaxy is
profound. `Red and dead' galaxies can't always have been dead because they
contain large masses of stars, and so at some point in the past they must have
been among the star-forming galaxies. There must therefore have been some physical process that converted
a star-forming galaxy on the Main Sequence into a red-and-dead galaxy, and this process
must have quenched the star-formation quickly (at least quickly
relative to the age of the Universe) to explain the relative dearth of galaxies in the green valley.
This conclusion has also produced a large literature and there is no consensus
about the
identity of this quenching process. But
the possibilities that have been suggested
include galaxy merging (Toomre 1977) with the merger scrambling the galaxy's velocity field
(turning it into an ETG) and the starburst triggered by the merger rapidly consuming
the gas; the expulsion of gas by a wind from an AGN (Cicone et al. 2014);
the infall of star-forming clumps into the centre of a galaxy 
which creates a bulge (and thus an ETG), which then reduces the star-formation rate
by stabilizing the disk
(Noguchi 1999; Bournaud et al. 2007; Martig et al. 2009; Genzel et al. 2011, 2014); plus
a plethora of environmental processes which either reduce the rate at
which gas is supplied to a galaxy or drive out most of the existing gas
(Boselli and Gavazzi 2006).

Almost all the work referenced above was based on galaxies discovered in optical surveys, even if
observations in other wavebands were used to estimate star-formation rates and stellar
masses. The launch of the {\it Herschel Space Observatory} (Pilbratt et al. 2010), which
observed the Universe in the wavelength range 70-600 $\mu$m,
gave astronomers the opportunity to look at the galaxy population
in a radically different way. 
Whereas Malmquist bias biases optical surveys
towards optically-luminous galaxies, and thus against dwarfs and
gas-rich galaxies, Malmquist bias
in the submillimetre waveband biases
surveys towards galaxies with large masses of interstellar dust.
One example of the different galaxyscape revealed by a
submillimetre survey
was the discovery 
of a new class of galaxy with large amounts
of cold dust but mysteriously very blue colours, which generally imply
an intense interstellar radiation field and much warmer dust
(Clark et al. 2015; Dunne et al. 2018).

In this paper, the third in a series of four
papers, we explore the new galaxyscape revealed by {\it Herschel}.
In the first paper (Eales et al. 2017 - E17), we used
the {\it Herschel} Reference Survey (HRS)  to investigate the assumption that
there are two separate classes of galaxy. The HRS was not actually selected
in the {\it Herschel} waveband, but was instead selected in the near-infrared
to be a volume-limited sample of nearby galaxies which would then
be observed with
{\it Herschel}. One great advantage of the HRS
is that the way it was selected means that it contains
most of the stellar mass in a local volume of space, and is thus
an excellent representation of the galaxy population {\it after} 12 billion
years of galaxy eviolution.  
When we plotted the positions of the HRS galaxies
in a diagram of SSFR versus stellar mass (left-hand panel of Figure 1),
we found that rather than the galaxies forming a separate
star-forming Main Sequence and a region of `red-and-dead' galaxies, they
actually fall on a single, curved Galaxy Sequence (GS), with the
morphologies of the galaxies gradually changing along the GS rather than
there being an obvious break between LTGs and ETGs.

\begin{figure*}
\includegraphics[width=140mm]{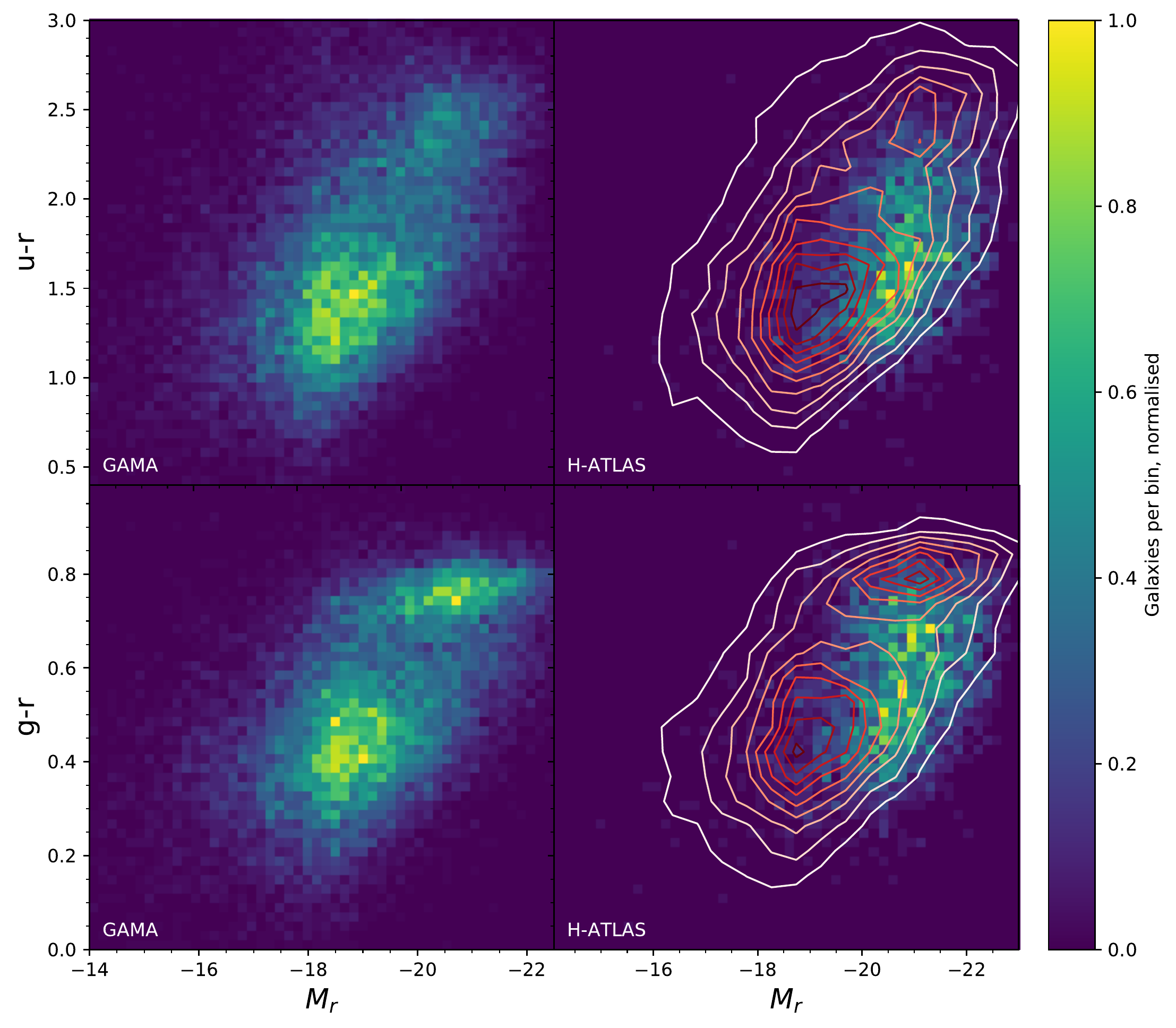}
  \caption{The distribution of galaxies
in the colour versus absolute $r$-band magnitude plane, with the colour
(see colour scale to the right) showing the density of
galaxies in this diagram.
The left-hand panels show the GAMA sample and the right-hand
panels show the H-ATLAS sample,
with the top panels showing $u-r$ colour versus absolute magnitude
and the bottom
panels showing $g-r$ versus abolute magnitude. 
The contours in the right-hand panels show the distributions for
the GAMA galaxies that are shown by the colour scale in the left-hand panels.
}
\end{figure*}

\begin{figure*}
\includegraphics[width=140mm]{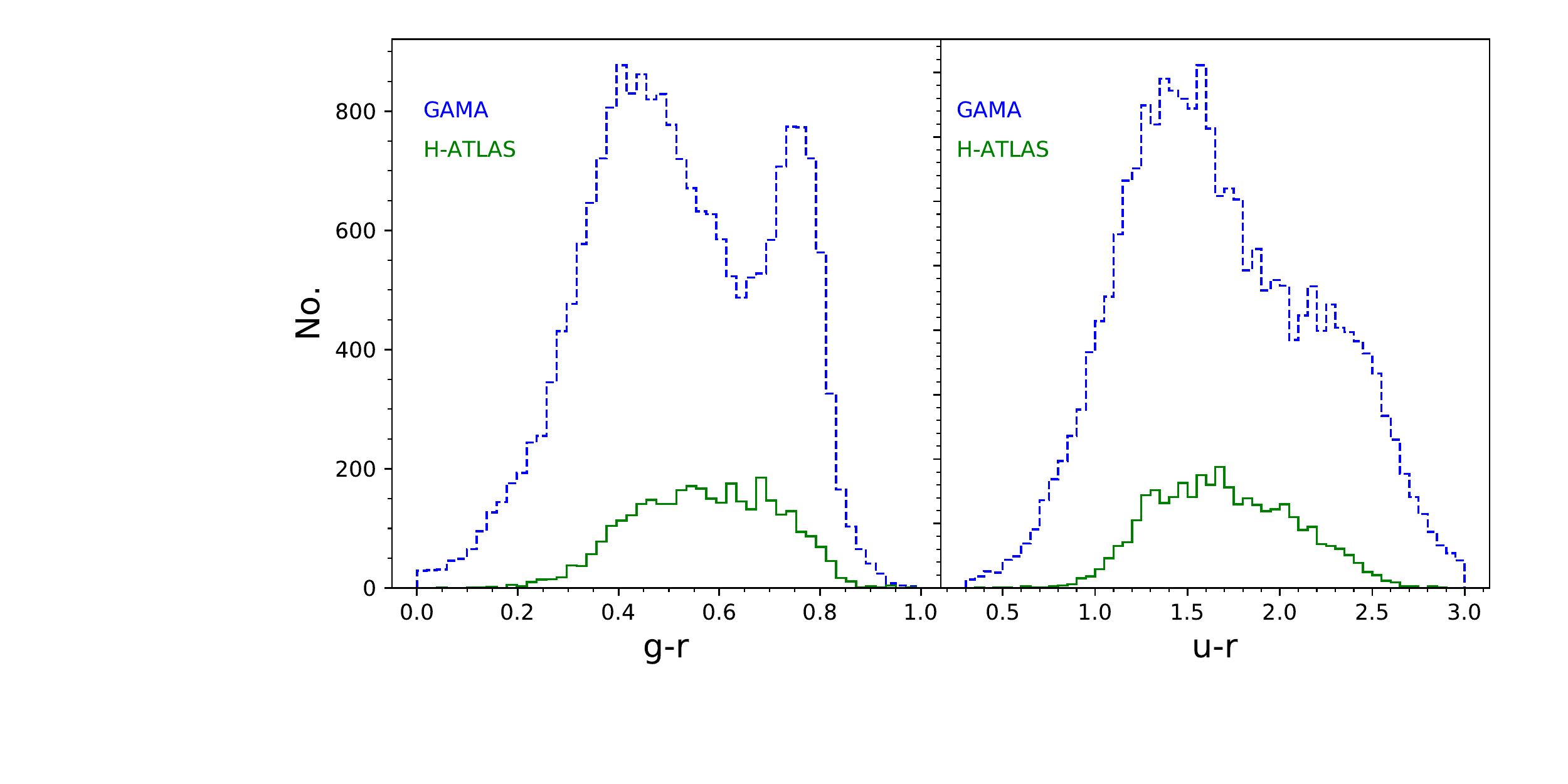}
  \caption{The distributions of $g-r$ colour (left) and
$u-r$ (right). The blue (dashed) histogram shows the
distribution for the GAMA sample and
the green (solid line) histogram the distribution for the sample
from H-ATLAS.
}
\end{figure*}

Our conclusion that the galaxy population is better thought of as a single
population rather than two physically-distinct classes agreed well with other recent
results. First, other groups have also found
evidence that the GS is curved, whether only star-forming
galaxies are plotted (Whitaker et al. 2014; Lee et al. 2015; Schreiber et al.
2016; Tomczak et al. 2016) or all galaxies are plotted
(Gavazzi et al. 2015; Oemler et al. 2017); we showed in E17 
that all recent attempts to plot the distribution of galaxies
in the Universe today
in the SSFR-versus-stellar-mass diagram are consistent once allowance
is made for selection effects
and for different ways of separating star-forming and passive galaxies.
Second, Pan et al. (2018) have shown that when the
H$\alpha$ line is used to estimate the star-formation rate
the apparent bimodal distribution
of galaxies in the SSFR-stellar-mass diagram 
is actually the result
of processes other than young stars producing H$\alpha$ emission.
Third, the ATLAS$^{\rm 3D}$ and SAMI integral-field spectroscopic surveys
of nearby galaxies also found no clear
kinematic distinction between ETGs and LTGs (Emsellem et al. 2011; Cappellari et al.
2013; Cortese et al. 2016).

Our second paper (Eales et al. 2018 - E18)
was based on a catalogue that was genuinely selected in the
{\it Herschel} waveband. 
The {\it Herschel} Astrophysical Terrahertz Large Area Survey
(H-ATLAS)
was the {\it Herschel} survey covering the largest area
of sky (660 deg$^2$) in five far-infrared and submillimetre bands (Eales et al. 2010).
All the data, including images, catalogues of sources and catalogues of the optical
and near-infrared counterparts 
is now public (Valiante et al. 2016; Bourne et al. 2016; Smith et al.
2017; Maddox et al. 2018; Furlanetto et al. 2018) and can be obtained
at \url{h-atlas.org}. In E18, we showed, first, that once a correction is made
for
Malmquist bias in the submillimetre waveband, 
the GS revealed by H-ATLAS is very similar to that found
in the volume-limited HRS. Second, we showed that previous attenpts to divide galaxies
into star-forming and passive galaxies missed an important population of red star-forming
galaxies. We showed that the space-density of this population is at least as great
as the space-density of the galaxies usually allowed membership of the star-forming
Main Sequence. Oemler et al. (2017) reached exactly the same conclusion, starting from
the SDSS galaxy sample and making a careful investigation of all the selection effects
associated with the SDSS.
Third, we found that galaxy evolution, investigated using several different methods,
is much faster at low redshifts than the predictions of the theoretical models, with
significant evolution by a redshift of 0.1 in the submillimetre luminosity
function (Dye et al. 2010; Wang et al. 2016),
the dust-mass function (Dunne et al. 2011) and
the star-formation-rate function (Marchetti et al. 2016; Hardcastle
et al. 2016).
We showed that the new results revealed by {\it Herschel} can be explained
quite naturally by a model without a catastrophic quenching process, in which most massive galaxies
are no longer being supplied by gas and in which the strong evolution and
the curvature of the GS are produced by the
gradual exhaustion of the remaining gas.

In this third paper of the series, 
we step back from trying to make broad inferences about galaxy evolution
and consider a detail, albeit an important one.
We first investigate where the H-ATLAS galaxies lie on the standard colour-versus-absolute
magnitude plots and whether this distribution differs from the
red-sequence/blue-cloud distribution seen for galaxies from
an optically-selected sample. We then
investigate whether the distributions of galaxies
in this diagram, whether produced from the SDSS or from {\it Herschel}, can be explained
by a continuous, curved Galaxy Sequence.

We assume a Hubble constant of 67.3 $\rm km\ s^{-1}\ Mpc^{-1}$
and the other {\it Planck} cosmological parameters (Planck Collaboration
2014).

\begin{figure}
\includegraphics[width=70mm]{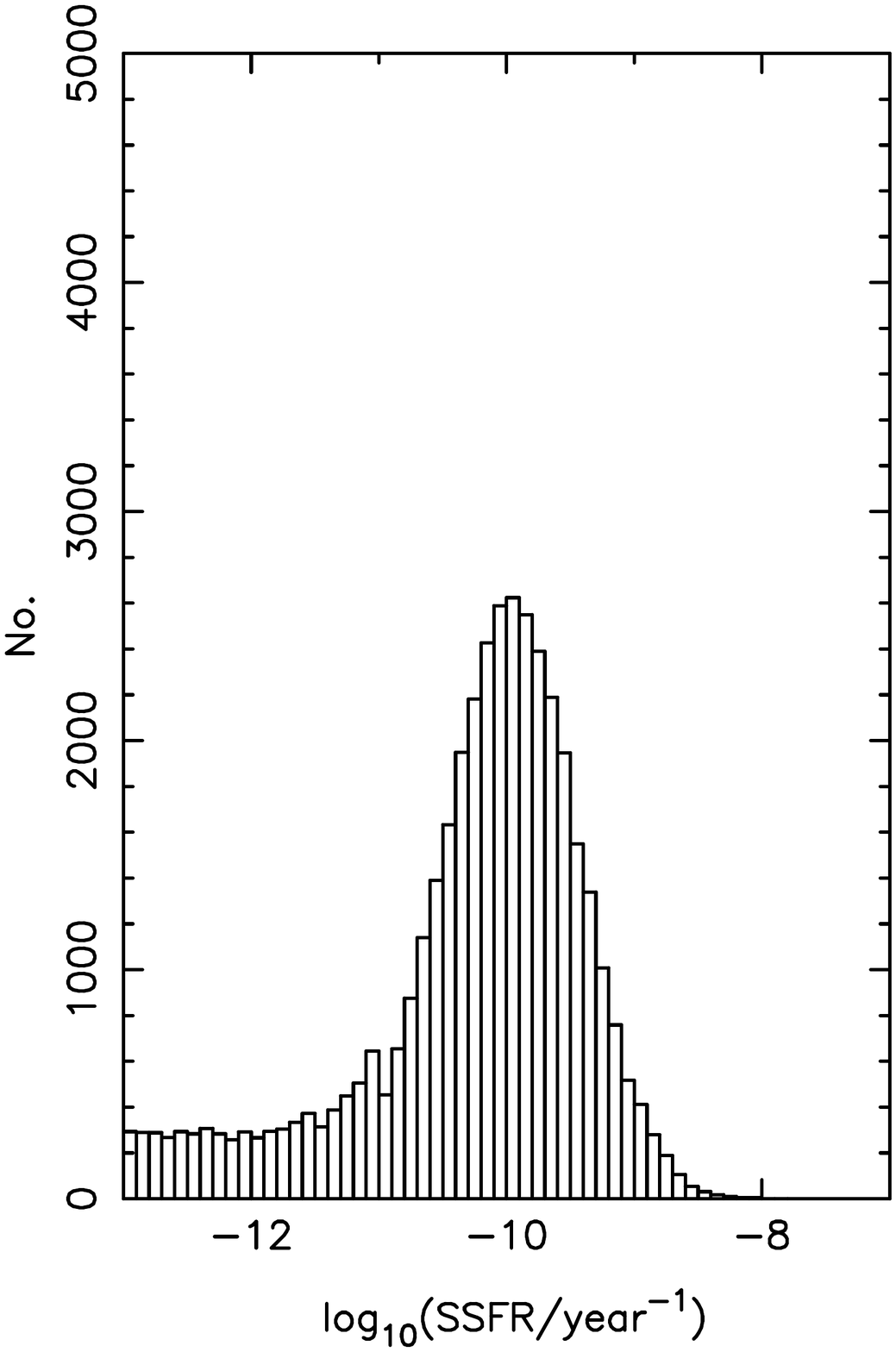}
  \caption{Histogram of specific star-formation rate
for the 40,000 galaxies generated in the Monte-Carlo
simulation described in \S 4.
}
\end{figure}

\begin{figure}
\includegraphics[width=70mm]{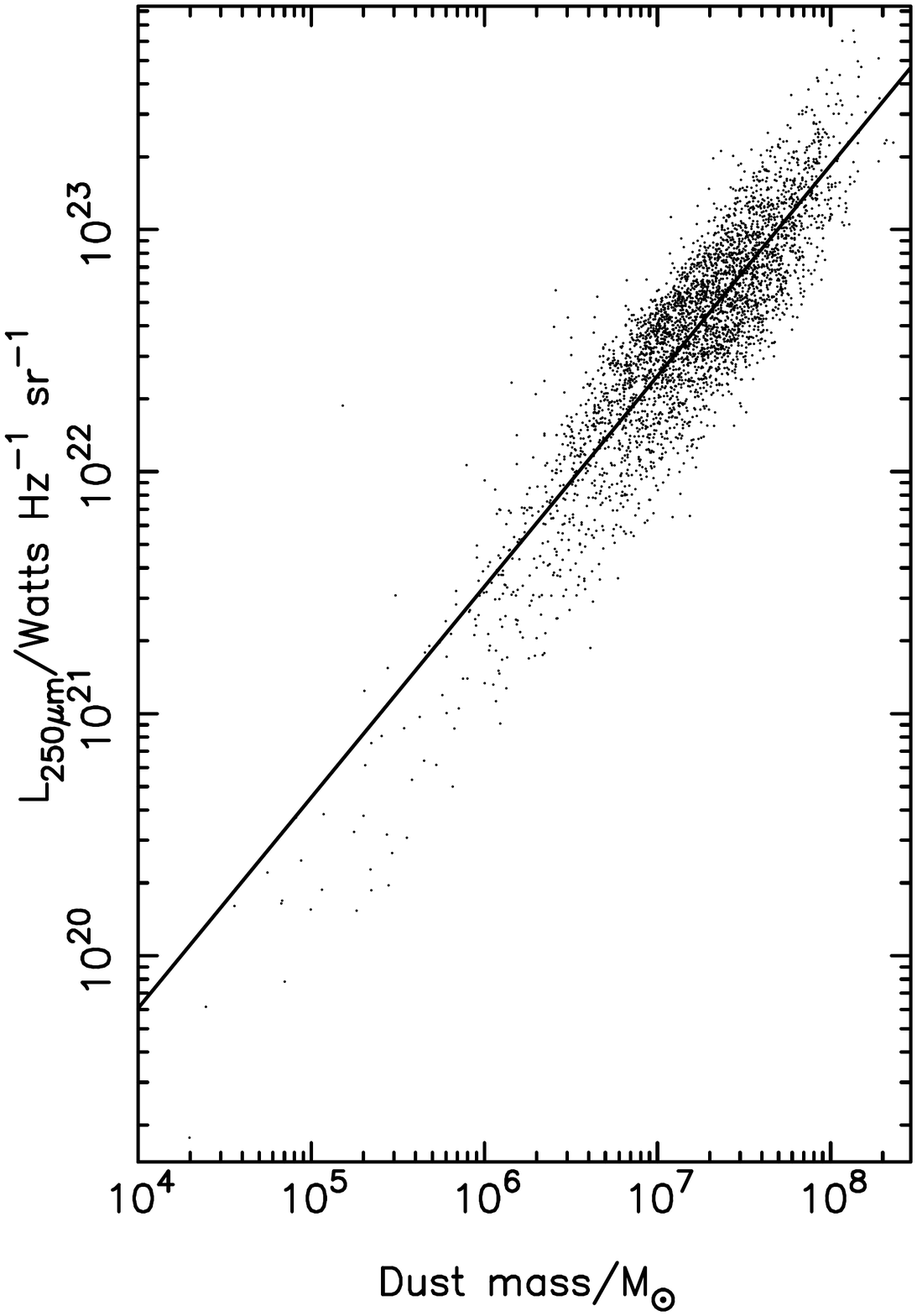}
  \caption{Luminosity at 250 $\mu$m versus dust mass for the
GAMA galaxies detected with {\it Herschel}. The dust masses
are the estimates from MAGPHYS (Driver et al. 2018).
}
\end{figure}

\section{The Samples}

For our optical sample we started from the galaxies detected in the Galaxy and Mass Assembly
project (henceforth GAMA). GAMA was a deep spectroscopic survey (Driver et al.
2009; Liske et al. 2015) complemented with matched-aperture
photometry throughout the $UV$, optical and IR wavebands (Driver et al. 2016).
We used the data from GAMA II (Liske et al. 2015),
which has a limiting Petrosian magnitude of $r < 19.8$.
We used the data from the
GAMA9, GAMA12 and
GAMA15 fields.
The optical sample we use here consists of all the galaxies in these
fields with 
reliable spectroscopic redshifts $<$0.1 and
consists of 20,884 galaxies. 

A crucial data-product from the GAMA survey that we use in our analysis
are estimates of the star-formation rates, stellar masses and dust masses.
These were obtained by the GAMA team (Driver et
al. 2018) by applying the MAGPHYS modelling
programme (Da Cunha, Charlot and Elbaz 2008) to the
matched-aperture photometry for each
galaxy, which extends from the $UV$ to the
far-infrared waveband (Wright et al. 2016).
Very briefly (see Driver et al. for a longer description
or the original paper for the full description), MAGPHYS combines
50,000 possible models of the stellar population of a galaxy with
50,000 models of dust in the ISM, balancing the energy
absorbed by the dust in the optical and $UV$ bands
with the
energy emitted by the dust in the far infrared, and matches the resultant
spectral energy distribution to the observed photometry.
Two of the merits of MAGPHYS are that it makes full use
of all the data and that it provides errors
on all the parameter estimates.
All tests that have been done of the MAGPHYS estimates (see E17
and E18) 
suggest that these estimates are robust.

For our submillimetre sample we started from H-ATLAS.
We used the same fields (GAMA9, 12 and 15) that were
used to produce the optical sample.
The 4$\sigma$ flux limit of the H-ATLAS survey in these fields is $\simeq$30 mJy
at 250 $\mu$m, the most sensitive wavelength, and there are 113,955 sources
above this flux limit (Valiante et al. 2016).
The H-ATLAS team used the SDSS $r$-band images to look for the optical counterparts to these
sources, finding 44,835 probable counterparts (Bourne et al. 2016). As our submillimetre sample,
we used the 3,356 sources in these fields with counterparts with spectroscopic redshifts
$<$0.1. Bourne et al. (2016) 
estimate that the H-ATLAS counterpart-finding procedure should have found
the counterparts to 91.3\% of the H-ATLAS sources with $z<0.1$. Virtually all
of these galaxies should have spectroscopic redshifts (Eales et al. 2018). 
Our sample should therefore be a close-to-complete submillimetre-selected sample
of galaxies in the nearby Universe.

Both samples are flux-limited samples, one in the optical waveband and
one in the submillimetre waveband.
Both will therefore be subject to Malmquist bias. In the optical sample,
optically-luminous galaxies will be over-represented because it is possible
to see these to greater distances before they fall below the flux limit
of the sample. In the submillimetre sample, galaxies with high submillimetre
luminosities will be over-represented for the same reason.

\begin{figure*}
\includegraphics[width=140mm]{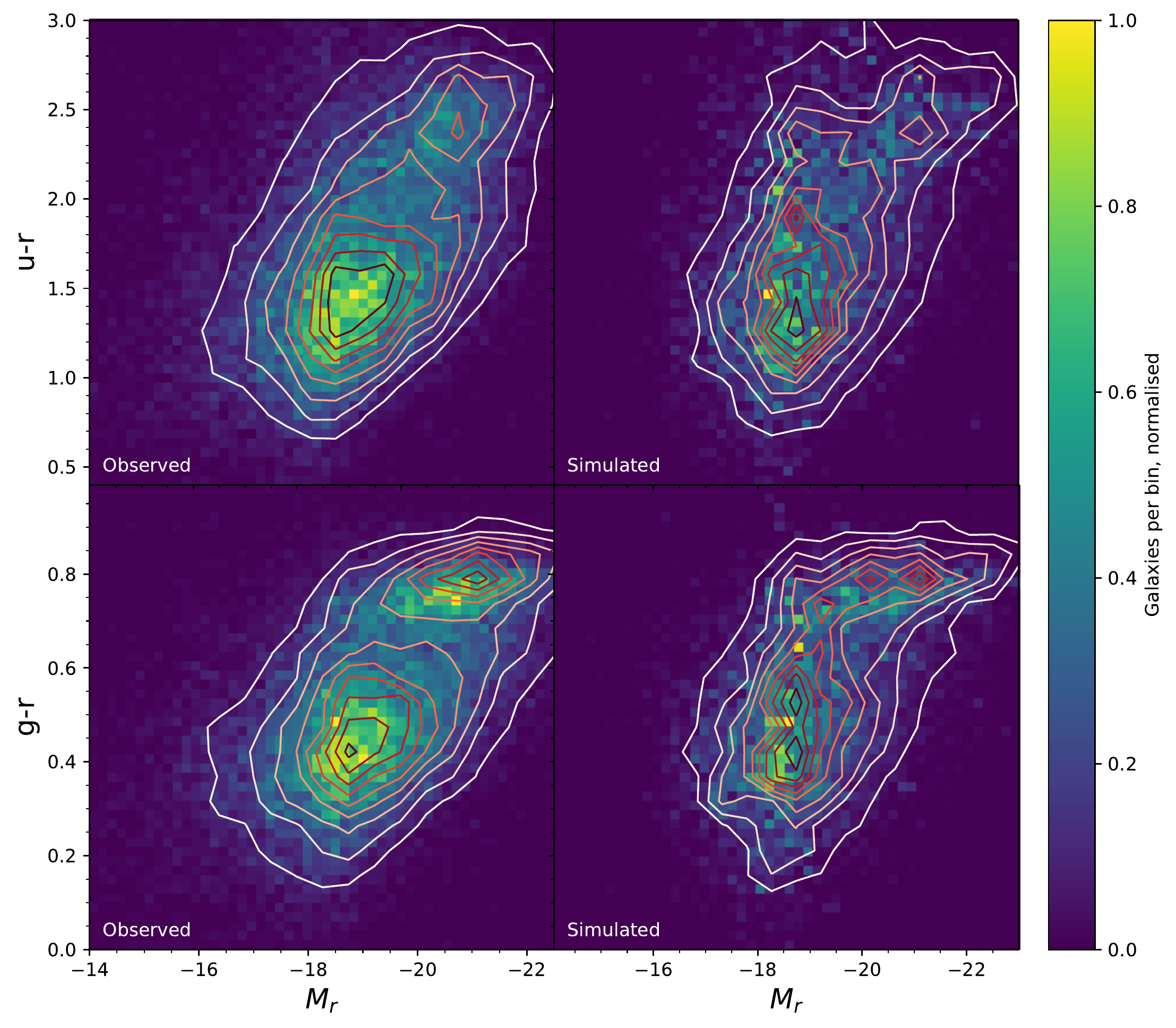}
  \caption{The distribution of galaxies
in the colour versus $r$-band absolute magnitude
diagram for the GAMA sample, with the
colour showing the density of galaxies in the
diagram (see colour bar to the right).
The left-hand panels show the observed
distributions, which are the same as
shown in the left-hand panels of Figure 2. 
The right-hand panels show the results of our
simulation of where the GAMA galaxies are expected to lie
in this digram if galaxies lie on
a continuous, curved Galaxy Sequence (\S 4).
}
\end{figure*}

\section{The Observed Colour Distributions}

For both samples, we calculated rest-frame $g-r$ and $u-r$ colours
and absolute $r$-band magnitudes for each galaxy using the GAMA matched-aperture 
Petrosian magnitudes
(Driver et al. 2016). In calculating the absolute magnitudes,
we used the individual k-corrections
for each galaxy calculated by the GAMA team (Loveday et al. 2012).

Figure 2 shows the distributions of both samples over the colour-absolute-magnitude
plane. 
The left-hand panels show the blue cloud and red sequence characteristic
of optical samples.
Both panels look very similar to the colour-absolute-magnitude
diagrams in Baldry et al. (2012 - their Figure 14), which is not surprising because
the two samples are both taken from the GAMA survey and are very similar in the
way they were selected.

The right-hand panels show the distributions of the H-ATLAS galaxies.
These distributions are quite different from those
for the optically-selected galaxies,
with the peak of the distribution for the submillimetre-selected galaxies shifted in both
absolute magnitude and colour from the blue
cloud. The colour offset is seen best in the colour distributions in
Figure 3. In the left-hand panel, which shows the histograms of $g-r$ colour, the peak of the distribution
for the submillimetre-selected galaxies is in the middle of the green valley. 
Thus in the diagram of $g-r$ colour versus absolute magnitude, the submillimetre-selected
galaxies
form a `green mountain' rather than the red sequence and blue cloud seen for
optically-selected galaxies.
The contrast between the optically-selected and submillimetre-selected galaxies
is not quite as dramatic for $u-r$ colour, since the peak seen for the
submillimetre-selected galaxies is not quite as well centred on the green valley (right-hand
panel of Figure 3).
However, the distributions for the optically-selected and submillimetre-selected
samples are still very different, with the distribution for the submillimetre-selected galaxies
having a single peak, shifted from the blue cloud
to brighter absolute magnitudes and in colour towards the centre
of the green valley. 

\section{Modelling the Distributions}

In this section, we test whether the difference
in the colour-absolute-magnitude distributions
is caused by the different selection effects for an optically-selected and 
submillimetre-selected sample. Based on the conclusion from our previous two papers (E17, E18),
we start with the assumption that in a diagram of SSFR versus
stellar mass, galaxies follow a single, continuous GS rather than a star-forming
Galaxy Main Sequence and a separate region of passive galaxies. Therefore, in the modelling
in this section, we are also testing whether the idea that there is a single Galaxy Sequence
is compatible with these very distinctive distributions in the
colour-absolute-magnitide diagrams.

\begin{figure*}
\includegraphics[width=140mm]{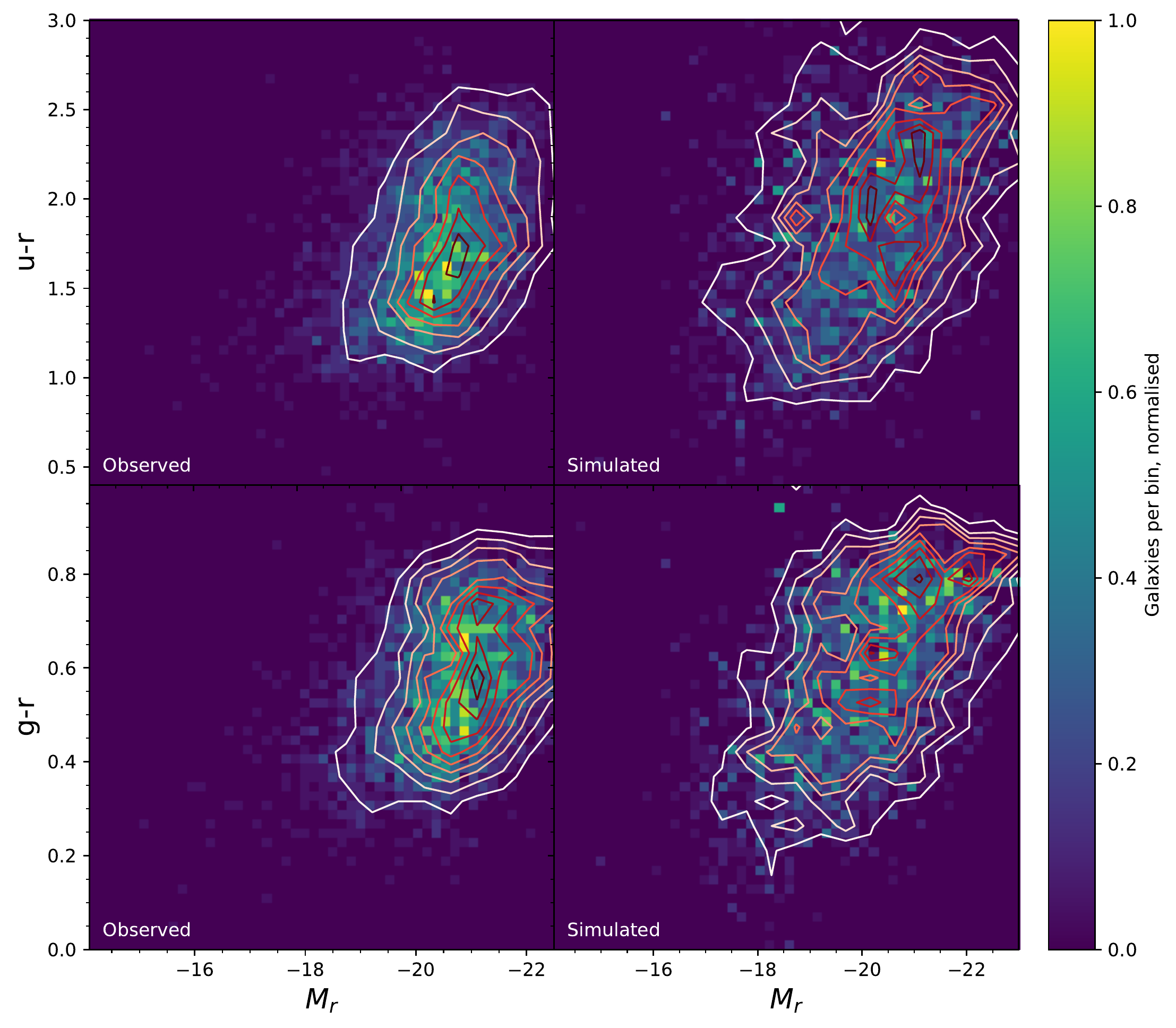}
  \caption{The distribution of galaxies
in the colour versus $r$-band absolute magnitude
diagram for the H-ATLAS sample, with the
colour showing the density of galaxies in the
diagram (see colour bar to the right).
The left-hand panels show the observed
distributions, which are the same as
shown in the right-hand panels of Figure 2.
The right-hand panels show the results of our
simulation of where the H-ATLAS galaxies are expected to lie
in this diagram if galaxies lie on
a continuous, curved Galaxy Sequence (\S 4).
}
\end{figure*}

Before we get into detail, we give a brief overview of our method.
There are three stages. In the first stage we use the GS to generate
values of SSFR and stellar mass for an artificial sample of galaxies with redshifts < 0.1.
The second stage is to associate an optical absolute magnitude and colour and a submillimetre
luminosity with each artificial galaxy. We do this using the GAMA sample,
by finding the GAMA galaxy closest to the artificial galaxy
in the SSFR-stellar mass plane,
and then assigning the real galaxy's absolute magnitude, colour and submillimetre
luminosity to the artificial galaxy. In the third stage we add the 
selection effects, finding the subsets of the artificial sample
that have flux densities brighter than the optical flux limit (an r magnitude
of 19.8) or the submillimetre flux limit (a flux density of 30
mJy at 250 $\mu$m).

\begin{figure*}
\includegraphics[width=140mm]{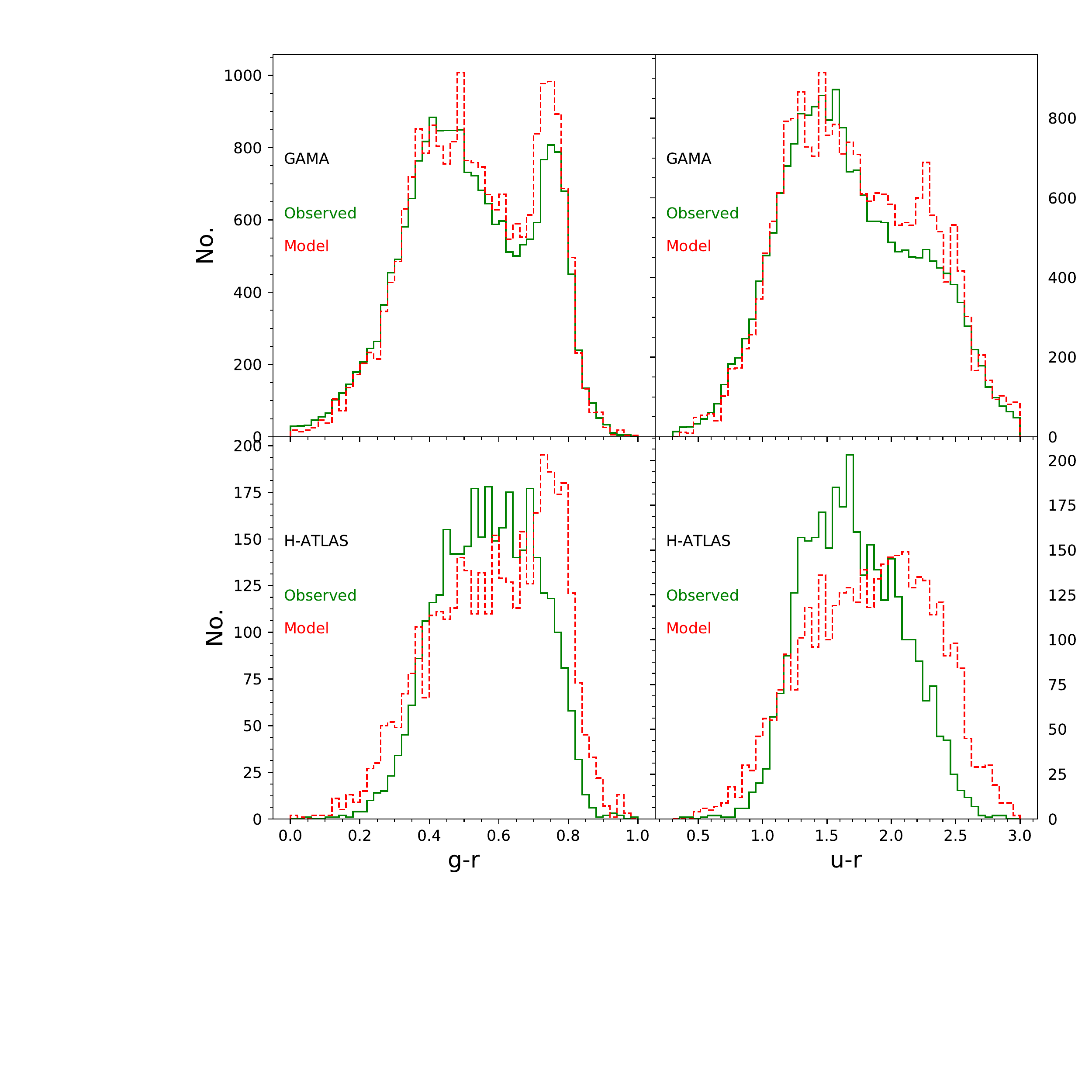}
  \caption{Histograms of colour for the GAMA and H-ATLAS samples.
In each panel the green solid line shows the histogram for the real sample
and the red dashed line shows the histogram predicted by our simulation
based on the assumption of a continuous, curved Galaxy Sequence.
The left-hand panels show $g-r$ colour and the right-hand panels
$u-r$ colour.
}
\end{figure*}

The aim of the first stage is to
create an artificial GS that resembles 
the empirical GS. 
The left-hand panel in Figure 1 shows the empirical GS from the {\it Herschel} Reference
Survey (E17). The HRS provides a complete inventory of LTGs down to a stellar
mass of $\rm \simeq 8 \times 10^8\ M_{\odot}$ and of ETGs down
to a stellar mass of 
$\rm \simeq 2 \times 10^{10}\ M_{\odot}$ (E17). It therefore misses low-mass ETGs,
which will fall in the bottom left-hand corner of the figure. However, the total stellar
mass of these missing
galaxies is quite small; E17 estimate that $\simeq$90\% of the total stellar
mass in ETGs with stellar masses $\rm >10^8\ M_{\odot}$ is contained in the
galaxies actually detected in the HRS. 
The figure shows that the morphologies of the galaxies 
gradually change
as one moves from the top-left to the bottom-right of the diagram.
E17 therefore concluded that galaxies follow a single, curved GS, with
galaxy morphology gradually varying along it, rather than there being
two separate distributions
of star-forming and passive galaxies. The small clump of ETGs at the bottom of the diagram is
almost certainly not significant because estimates of SSFR from SED-fitting programs,
such as MAGPHYS and CIGALE, for galaxies with $\rm log_{10}(SSFR)<-12.0$
are extremely unreliable because the shape of the SED depends so weakly on SSFR
in this part of the diagram (Hunt et al. in preparation).

We generated 
an artificial sample in the following way using a GS that resembles the HRS GS.
The details of our method, which are described in the next
two paragraphs, are a little complex, and since the details are unimportant as long
as the method does produce a continuous GS, which we show below that it does, the reader
may wish to skip over these paragraphs.
The first step was to use a random-number generator to
generate stellar masses using the
stellar mass functions for star-forming
galaxies and passive galaxies (Baldry et al. 2012) as probability distributions.
Baldry et al. divided galaxies into `passive' and `star forming' using
a line on the colour-absolute-magnitude diagram chosen to separate the
red sequence and the blue cloud. We made the approximation that this
division is equivalent to a constant value of SSFR, $SSFR_{cut}$,
and used the stellar mass function
for star-forming galaxies to generate masses for artificial galaxies for $SSFR > SSFR_{cut}$
and the stellar mass function for passive galaxies to generate
masses for artificial galaxies for $SSFR < SSFR_{cut}$. Note that this is a fairly
crude approximation, and E18 showed that galaxies in the red-sequence
part of the colour-absolute-magnitude diagram extend to quite high values
of SSFR (see their Figure 5). Given the crudeness of this approximation, we needed a way
to tune our model. One possible way would have been to vary the value
of $SSFR_{cut}$. Instead, for reasons of practicality, we chose
a fixed value of $SSFR_{cut}$, but used a probability distribution
for the stellar masses for galaxies with $SSFR < SSFR_{cut}$ equal
to the stellar mass function for passive galaxies multiplied by
the parameter $f_{passive}$, the one free parameter in our model.
In detail, we chose a value for $SSFR_{cut}$ of $\rm 10^{-11}\ year^{-1}$
and used a lower mass limit of
$\rm 10^8\ M_{\odot}$. We adjusted $f_{passive}$ until
we got roughly the right number of galaxies on the red sequence; we found
we got reasonable agreement with
a value of 0.5. We discuss the effect of varying this
parameter in \S 6.

We assigned a value of SSFR to each galaxy in a different
way depending on whether the galaxy had been assigned to
the region below or above $SSFR_{cut}$.
We assumed that the galaxies with $SSFR < SSFR_{cut}$
are uniformly
spread over the range range $\rm -13.0 < log_{10}(SSFR) < -11.0$, using a random-number generator
to generate a value for the SSFR of each galaxy.
We generated values of SSFR for galaxies with $SSFR>SSFR_{cut}$ in the
following way.
The upper part of the empirical GS 
is well fit by a second-order polynomial (Figure 1, left panel).
We used this polynomial to generate values of SSFR for galaxies
with $SSFR>SSFR_{cut}$ on the assumption
that
star-forming galaxies have a gaussian distribution in $log_{10}(SSFR)$ around this line
with $\sigma=0.5$, a value chosen to match the observed spread of the HRS
galaxies.

\begin{figure*}
\includegraphics[width=140mm]{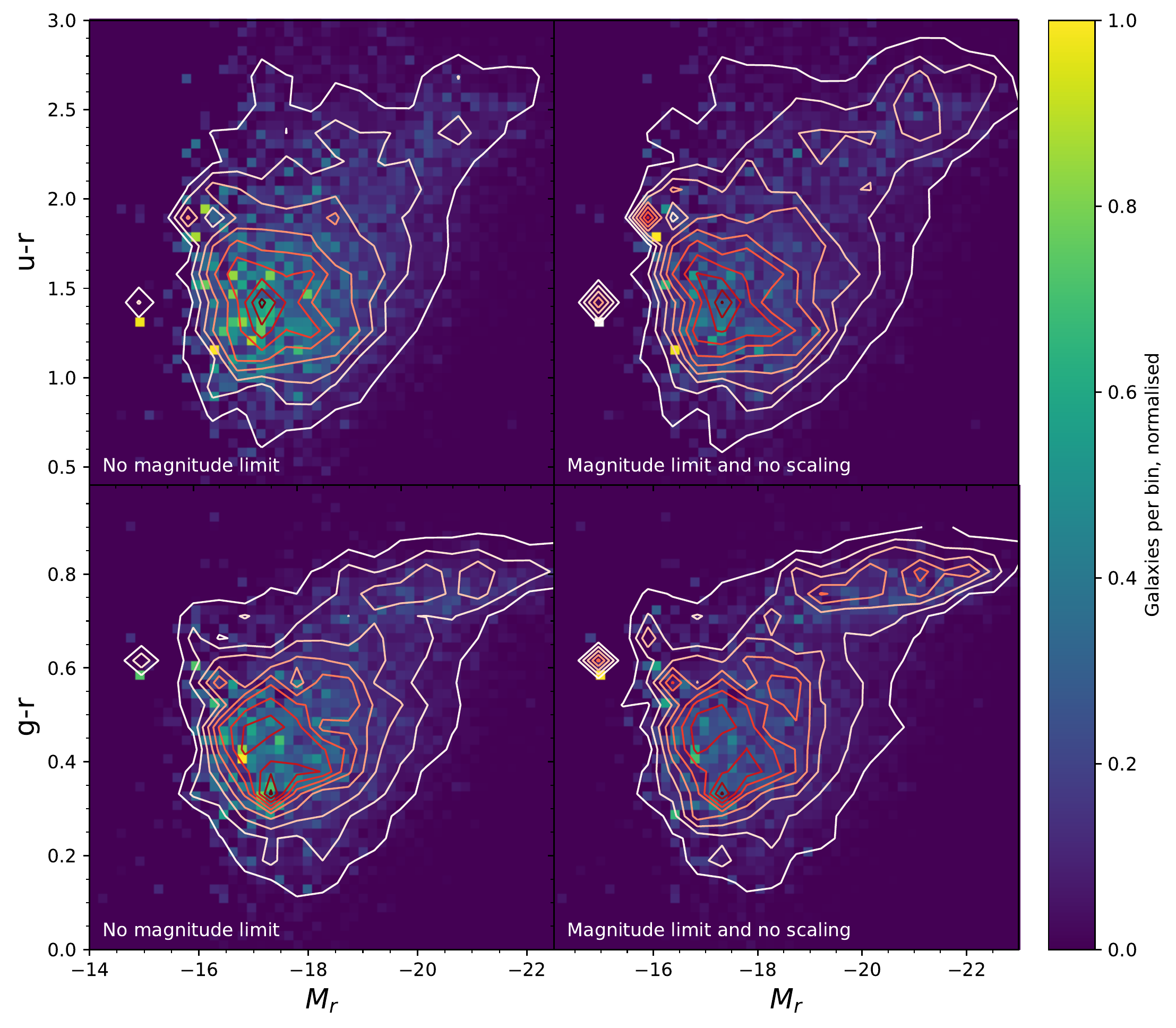}
  \caption{The distribution of galaxies
in the colour-versus-absolute-magnitude
diagram, with the colour showing the density of
galaxies in this diagram (see colour scale to the right).
The left-hand panels show the distribution for all 40,000
galaxies in our artificial sample (\S 4) when no flux limit is
imposed on the sample. The right-hand panels show the result
for an artificial sample with $f_{pass}$,
the multiplicative
factor for the stellar mass function of passive galaxies (\S 4),
set to 1, for the artificial galaxies that satisfy the GAMA
magnitude limit ($r < 19.8$). To see the effect of
changing the value of this parameter, these panels should
be compared with the right-hand panels for Figure 6, for which
$f_{pass} = 0.5$.
}
\end{figure*}

Using this method, we generated values
of SSFR and stellar mass for 40,000 galaxies, with the
results being shown in the right-hand panel of Figure 1. 
This procedure generates a smooth, curved GS that looks
reasonably like the empirical GS in the left-hand panel.
Figure 4 shows a histogram of SSFR for the sample, confirming that we have
generated a sample of galaxies with a smooth distribution in SSFR rather than
the bimodal distribution characteristic of a separate star-forming Main Sequence
and region of passive galaxies.

There are two important points we wish to emphasise here.
The first point  is that we do not claim that the Galaxy Sequence we
are using is a perfect representation of the true Galaxy Sequence.
The form of this is still poorly known, mainly because of the
difficulty of measuring star-formation rates in ETGs (E17; Pan et al. 2018).
The aim of our modelling is the more modest one of testing whether a 
continuous Galaxy Sequence that at least looks somewhat like the
empirical Galaxy Sequence shown in the left-hand panel
of Figure 1 can produce the very distinctive distributions
in the colour-absolute-magnitude diagrams, or whether these distributions can
only be produced by galaxies lying in two separate regions of the
SSFR-versus-stellar-mass diagram e.g. a star-forming Main Sequence
and a separate region of passive galaxies.

The second important point is that our division of galaxies
into ETGs and LTGs in the first part of our analysis is not
part of a circular argument that
necessarily produces a red sequence and a blue cloud
in the colour-absolute-magnitude diagrams. 
We made this division as part of our procedure for generating
a continuous Galaxy Sequence and we did not use this division at any future
point in the modelling. The division of galaxies into two classes
in the first stage of the modelling could only lead to galaxies falling in two
separate regions of the colour-absolute-magnitude diagrams if it also led to
galaxies falling in two separate regions of the SSFR-stellar-mass diagrams.
Figure 1 (right panel) and Figure 4 confirm that this is not the case.
Since we have achieved our aim of generating a continuous Galaxy Sequence, the
division in to ETGs and LTGs and all the other details of the first stage of our
analysis
are now irrelevant and can now be thankfully forgotten.

As the final step of the first stage in the
modelling, we used a random-number generator to produce a redshift for each galaxy
on the assumption that our sample is composed of all the galaxies
in a particular region of sky out to a redshift of 0.1.

The second stage is to assign an absolute magnitude, colour and
submillimetre luminosity to each artificial galaxy.
We did this by using the GAMA sample effectively as a look-up table.
Each GAMA galaxy has an absolute magnitude, optical colours, and through the
MAGPHYS modelling (\S 2), estimates of SSFR, stellar mass and dust mass.
For the artificial galaxies with
$\rm log_{10}(SSFR) > -12.0$ we found the GAMA galaxy
that is closest to in the $\rm log_{10}(SSFR)$, $\rm log_{10}(M_*)$ space
to the artifical galaxy.
Lower values of SSFR have very large errors, and so for artificial galaxies
with values of SSFR below this limit, we found the GAMA galaxy
that also has
$\rm log_{10}(SSFR) < -12.0$ and that is closest in $\rm log_{10}(M_*)$
to the artificial galaxy. In both cases, we assigned the absolute magnitude, colours
and MAGPHYS dust mass of the GAMA galaxy to the artificial galaxy.

The one remaining step in this stage is to assign a submillimetre
luminosity to each artificial galaxy. We again did this using the GAMA results.
Figure 5 shows 250-$\mu$m luminosity plotted against the MAGPHYS dust-mass
estimate for all the GAMA galaxies that were detected with {\it Herschel}
at 250 $\mu$m.
The best-fitting line is given by:

\smallskip
$$
{\rm log_{10}(L_{250}/Watts\ Hz^{-1}\ sr^{-1}) = 0.87 log_{10}(M_d / M_{\odot}) + 16.3}
$$
\smallskip

\noindent with a dispersion in $\rm log_{10}(M_d)$ of 0.21. 
We used this relationship and the MAGPHYS dust mass associated
with each artificial galaxy to assign a 250-$\mu$m luminosity
to each artificial galaxy.

The final stage is to add the selection effects. This is quite easy since
each artificial galaxy has an $r$-band absolute magnitude, a 250-$\mu$m luminosity
and a redshift. It is thus simple to calculate the $r$-band magnitude and the 250-$\mu$m
flux density of each galaxy, and thus determine which galaxies would be brighter
than either the GAMA $r$-band magnitude limit (19.8) or the
the H-ATLAS 250-$\mu$m flux density limit (30 mJy)\footnote{In carrying
out the optical calculation we used the k-correction from the closest
GAMA galaxy in the SSFR-M$_*$ plane. In calculating the
250-$\mu$m flux density we assumed a dust temperature of 20 K.}

\section{Results}

Of the 40,000 galaxies in the artificial sample, 21,847  have $r$-band magnitudes $<$19.8,
the magnitude limit for GAMA, and 3,646 have 250-$\mu$m flux densities $\rm > 30\ mJy$,
the flux-density limit for H-ATLAS - both fairly similar to the numbers in
the real samples. 
Although the original number of artificial sources
was chosen to give roughly the correct number of sources in the real samples,
the fact that the ratio of the numbers in the artificial submillimetre and optical
samples is similar to the ratio of the numbers in the real samples
is
circumstantial
evidence in favour of our method.

\begin{figure}
\includegraphics[width=90mm]{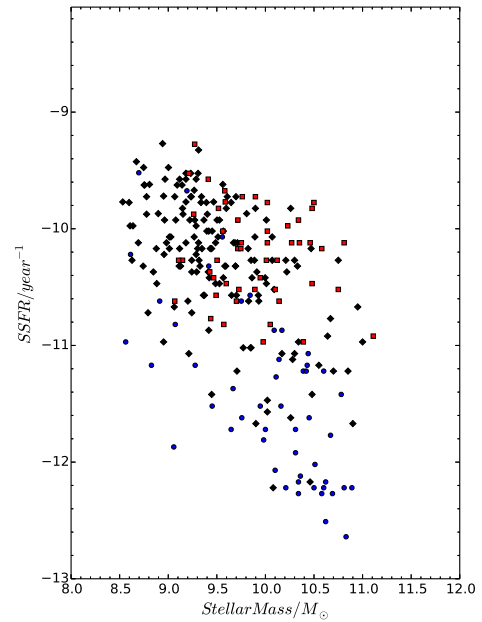}
  \caption{The distribution of the HRS galaxies in the SSFR-M$_*$ plane with the
points coloured to show the ratio of optical ($r$-band) to submillimetre (250 $\mu$m)
luminosity for each galaxy. The blue circles show the 50 galaxies with the
highest ratios of optical-to-submillimetre luminosity, the red squares show the 50 galaxies
with the lowest ratios of optical-to-submillimetre luminosity, and the
black diamonds the other HRS galaxies.
}
\end{figure}
The left-hand panels
of Figure 6 show again the real colour-absolute-magnitude distributions for the
GAMA sample and the right-hand panels show
the distributions for the galaxies in the optically-detected
sample of artificial galaxies.
Figure 7 shows the same diagram for the observed and artificial submillimetre-selected
samples. Figure 8 shows histograms of the colour for the real and artificial
optically-selected and submillimetre-selected samples.

Let us first consider the results for the optically-selected sample. Figures 6 and
8 show
that our simulation reproduces the qualitative features of
the distribution of the galaxies in the colour-versus-absolute-magnitude
diagram. It produces a clear red sequence and blue cloud, although it does not
produce the quantitative details of the distributions; in particular, there are slightly
too many galaxies in the red sequence. 

Now let us consider the results for the submillimetre-selected sample,
shown in Figures 7 and 8. Here the simulation again produces qualitatively
the
main feature of the observed distribution,
since it 
produces
a green mountain shifted in absolute magnitude and colour from
the blue cloud for the optically-selected galaxies. 
However, the quantitative agreement between the predictions and the
observations is less good than for the optical sample, since
there
are too many artificial galaxies on the red side of the 
green mountain.

\section{Discussion}

The comparison between the predictions and
the observations show that we have achieved
our main goal. We have shown that the distinctive features of
the distributions of optically-selected and submillimetre-selected
galaxies in colour-versus-absolute-magnitude diagrams can
be qualitatively produced 
from a single, continuous GS.

The reader may be concerned that our analysis is circular 
because of our use of the properties of the GAMA galaxies in the 
second part of the analyis (\S 4).
However, the only use
we have made of this sample is to map
the transformation between
the SSFR/M$_*$ space and absolute-magnitude-colour space; since the
GAMA sample provides good coverage of the former, there is no
reason to think that this procedure should introduce any biases.

Why does a single, continuous GS produce such very
different distributions for an optically-selected and
a submillimetre-selected sample?
We can understand this qualitatively in the following way. 

First, let us consider
the optically-selected sample. 
In the left-hand panels of Figure 9 we show the colour-absolute-magnitude
diagrams for the complete artificial sample of 40,000 galaxies without
requiring the galaxies to be brighter than the r=19.8 magnitude limit.
The red sequence and blue cloud are still apparent, showing the
red sequence and the blue cloud are {\it not} the consequence
of the magnitude limit but instead the result of the geometric mapping
of SSFR/M$_*$ to absolute magnitude/colour, which distorts a
continuous GS into the distinctive red sequence and blue cloud.
As we pointed out in E17, a red sequence is
naturally produced because all galaxies
with $\rm SSFR 
< 5 \times 10^{-12}\ year^{-1}$ have almost the same colour.

The right-hand panels of Figure 9 show the colour-absolute-magnitude
diagrams for the artificial galaxies but 
with the inclusion again of the $r=19.8$ magnitude limit and with $f_{passive}$,
the one free parameter in our model (\S 4), set to a value of 1.0 rather
than 0.5. The appearance of the panels is not very different from
the distributions for $f_{passive}=0.5$ shown in the right-hand
panels of Figure 6, but a plot of the colour histograms show
that increasing $f_{passive}$ to 1.0
leads to an increase in the number of galaxies on the red sequence,
which is already slightly too high for $f_{passive}=0.5$.

Although the implications of this result
are profound (see below), it is important to be clear about its
limitations.
We have
shown that the existence of the green valley in plots of colour versus absolute
magnitude is not evidence that there are two distinct classes of galaxy
(the same is true of plots of the strength of the 4000\AA\ break versus
absolute magnitude - E17), since a green valley is naturally produced
even if there is a single, continuous Galaxy Sequence in a plot
of the intrinsic quantitites: SSFR and stellar mass. However, we have not proved
there is only a single galaxy class
because a bimodal distribution of
galaxies in the plot of SSFR verses stellar mass would also naturally produce
a green valley. 
The importance of our result is that
the existence of the green valley has led to the conclusion that there
must be some rapid quenching process that moves a galaxy rapidly (in cosmic terms)
from the blue cloud to the red sequence (\S 1). Our demonstration
that the
green valley does not imply the existence of two separate distributions
of galaxies in the SSFR-stellar-mass diagram shows that the discovery of the
red sequence sixty years ago was something of a red herring.

Let us now turn to the submillimetre-selected sample. By stacking spectra, E18 showed
that the red colours of the galaxies in H-ATLAS are not the
result of dust reddening but of genuinely old stellar populations.
The fact that the green mountain has redder colours than the blue cloud
is therefore not the result of the {\it Herschel} galaxies containing
more dust.

The key diagram that shows the true cause of the green mountain
is Figure 10. In this diagram, we plot again the HRS galaxies
in the SSFR-M$_*$ diagram, this time colour-coding the points by the
ratio of optical-to-submillimetre luminosity for each galaxy.
We have calculated this ratio for each galaxy using the
$r$-band magnitudes in Cortese et al. (2012) and the
250-$\mu$m flux densities given in Ciesla et al. (2012).
For any flux-limited sample, the accessible volume for a galaxy
scales roughly as luminosity$^{1.5}$. Therefore, the ratio of the
optical-to-submillimetre luminosity is a measure of the relative Malmquist
bias in the two wavebands. For example, if the ratio was the same for every
galaxy, we would expect very similar distributions in the colour-versus-absolute-magnitude
diagram for an optically-selected and a submillimetre-selected sample.
In Figure 10, the blue points show the 50 galaxies with the
highest ratio of optical-to-submillimetre luminosity, the red points
the 50 galaxies with the lowest values of this ratio, and the black points
the galaxies with intermediate values.
The blue points are therefore the galaxies we would expect to see overrepresented
in an optically-selected
sample and underrepresented in a submillimetre-selected sample, and vice versa.
The blue points are mostly massive galaxies with low values of SSFR, which fall on
the red sequence in colour-absolute-magnitude
diagrams. So it
is not surprising that we see a red sequence for an optically-selected
sample but not for a submillimetre-selected sample. The red points are also fairly massive
galaxies but with much higher values of SSFR and are
at a position in the diagram where the GS is curving down. 
The galaxies in this part of the GS will therefore tend to be over-represented in a
submillimetre-selected sample.
It is therefore Malmquist bias boosting the representation of this part of the
GS that causes the distinctive green mountain in a submillimetre-selected sample.
This effect will also lead to these galaxies being underrepresented in an
optical sample, deepening the green valley.

We can also qualitatively explain why the difference in colour between the
{\it Herschel}-selected galaxies and the blue cloud is greater for
$g-r$ than for $u-r$ (Figure 3). The stacked spectra in E18 show that
the {\it Herschel}-selected galaxies have both a significant 4000\AA\ break,
indicating an old stellar population, and an upturn in the $UV$ part of
the spectrum, indicating large numbers of young stars.
The $g-r$ colours will be more sensitive to the old stellar population,
dragging the colour distribution for the {\it Herschel} galaxies
towards the red sequence; the $u-r$ colours will be more sensitive
to the $UV$ emission from young stars, pulling the colour distribution
towards the star-forming blue cloud.

Our result removes the most obvious evidence for rapid quenching
in the galaxy population. There are
a number of other recent studies, using a variety of methods, that 
have also investigated 
whether rapid quenching is an important part of
galaxy evolution.
Casado et al. (2015) used a comparison of estimates
of the recent star-formation rate in galaxies with star-formation estimates for
the same galaxies
at an earlier time to show that there 
is no evidence of
a rapid quenching process in low-density evironmeents, although there is
evidence for this
in dense environments. 
Using a similar technique Schawinski et al. (2014)
found evidence for slow quenching in LTGs but rapid quenching
in ETGs.
Peng, Maiolino and Cochrane (2015)
used the metallicity distributions of star-forming and passive
galaxies to argue that the evolution from one population to the other
must have occurred over $\simeq$4 billion years - slow 
quenching. Schreiber et al. (2016) concluded that at high redshift
the decrease in the slope of the star-forming Main Sequence at high stellar masses
can be explained by slow quenching.
Although the same problem of measuring accurate values for SSFR at low values
of SSFR exists at high redshift, the best evidence for galaxies lying in two distinct regions
of the SSFR-stellar-mass diagram comes from high-redshift samples
(Elbaz et al. 2017 - their figure 17; Magnelli et al. 2014 - their figure 2).

Finally, a couple of very recent studies have investigated
directly the evolution of the galaxies in the green-valley part
of the colour-absolute-magnitude diagram.
Bremer et al. (2018)
used the colours and structures of the galaxies in the green valley to
show that galaxies are evolving slowly across this region. 
A follow-up study by Kelvin et al. (2018) has used the structures of the green-valley galaxies
to show that the transition across this region does not require a rapid
quenching process. 
Our results nicely explain these two results, since if the green valley is
an artefact of the geometric transformation from intrinsic to observed
quantities, there is no reason to expect galaxies
in this part of the diagram to be evolving quickly.

One possible way to reconcile all of the results above, as we argued in E17, would be
if most galaxies evolve by gradual processes (slow quenching) with the 14\% of
ETGs that are slow rotators (Cappellari et al. 2013) being produced by a rapid-quenching process.

\section{Conclusions}

We have shown that the galaxies found in
a sample selected at submillimetre wavelengths have a very different
distribution in a colour-absolute-magnitude diagram 
than galaxies selected at optical wavelengths. Whereas
the optically-selected galaxies form a red sequence and a blue cloud with
a green valley in between, a submillimetre-selected sample forms
a green mountain.

We have shown that we can reproduce the qualitative features of 
both distributions if
galaxies lie on
a single, curved, continuous Galaxy Sequence in a plot of SSFR versus stellar mass.
The cause of the red sequence and the blue cloud is the
geometric mapping between stellar mass/specific star-formation rate and absolute magnitude/colour,
which distorts a continuous Galaxy Sequence in the diagram of intrinsic
properties into a bimodal distribution in the diagram of observed
properties. The cause of the green mountain is Malmquist bias in the
submillimetre waveband, leading galaxies on the curved part of the Galaxy Sequence
to be overrepresented in a submillimetre survey. This effect, working in reverse, causes
galaxies on this part of the Galaxy Sequence to be underrepresented in an optical
sample, deepening the green valley.

The existence of the green valley for an optical sample is therefore not evidence
(1) for there being two distinct populations of galaxies, (2) for galaxies in this region evolving
more rapidly than those in the blue cloud and red sequence, (3)
for rapid quenching processes in the
galaxy population.

\section*{Acknowledgements}

We thank the many scientists who have contributed to
the success of the {\it Herschel} ATLAS and the {\it Herschel} Reference
Survey.
We are also grateful to Leslie Hunt for useful conversations about the relative strengths and
limitations of the SED-fitting programs MAGPHYS, CIGALE and GRASIL and for
making availble a draft of her paper on the subject.
This research made use of Astropy, a community-developed core Python package for Astronomy (Astropy Collaboration
2013), and Matplotlib, a Python 2D plotting library (Hunter 2007).
EV and SAE acknowledge funding
from the UK Science and Technology Facilities Council consolidated grant ST/K000926/1.
MS and SAE have received
funding from the European Union Seventh Framework Programme  ([FP7/2007-2013]
[FP7/2007-2011])  under  grant
agreement No. 607254.
LD and SM acknowledge support from the European Research Council (ERC)
in the form of Consolidator Grant {\sc CosmicDust} (ERC-2014-CoG-647939, PI H\,L\,Gomez).
LD, SJM and RJI acknowledge
support from the ERC in the form of the Advanced Investigator Program,
COSMICISM
(ERC-2012-ADG
20120216,
PI  R.J.Ivison).
M.J.M.~acknowledges the support of the National Science Centre, Poland
through the POLONEZ grant 2015/19/P/ST9/04010.






\bsp	
\label{lastpage}
\end{document}